# Moral intuitions behind deepfake-related discussions in Reddit communities

**Abstract**

Deepfakes are AI-synthesized content that are becoming popular on many social media platforms, meaning the use of deepfakes is increasing in society, regardless of its societal implications. Its implications are harmful if the moral intuitions behind deepfakes are problematic; thus, it is important to explore how the moral intuitions behind deepfakes unfold in communities at scale. However, understanding perceived moral viewpoints unfolding in digital contexts is challenging, due to the complexities in conversations. In this research, we demonstrate how Moral Foundations Theory (MFT) can be used as a lens through which to operationalize moral viewpoints in discussions about deepfakes on Reddit communities. Using the extended Moral Foundations Dictionary (eMFD), we measured the strengths of moral intuition (moral loading) behind 101,869 Reddit posts. We present the discussions that unfolded on Reddit in 2018–2022 wherein intuitions behind some posts were found to be morally questionable to society. Our results may help platforms detect and take action against immoral activities related to deepfakes.

## 1 Introduction

Deepfakes are AI-based synthetic content; they are currently becoming popular on many social media platforms while simultaneously making headline news for their potential to lead to social harm. Deepfakes are typically double-edged swords, as they could potentially be used for deceptive purposes at the personal relationship or social levels. Thus, it is a matter of the moral intuition behind their use. In examinations of the origin of deepfakes in social media, it has been found that a subreddit on the Reddit platform named "r/deepfakes" was created to share pornographic videos of celebrities (van der Nagel, 2020). This indicates that the moral intuition behind such a community was to create and share pornographic videos of famous actresses. Due to Reddit's rule against non-consensual nude content and the pressure from the community, this subreddit was banned from the platform (Wik, 2022; Gamage et al., 2022).

However, since that subreddit ban in 2018, many discussions on deepfakes have occurred in many other Reddit communities. A recent study analyzed major topics emerging from discussions about deepfakes on the Reddit platform and revealed implications for society at large. Importantly, their findings raised alarming concerns for discussions relating to deepfakes (Gamage et al., 2022). It appeared that the community on the Reddit platform supports and encourages creating more deepfakes and even enabling a marketplace on deepfakes, regardless of its possible harmfulness (Gamage et al., 2022):example posts include *"I can do your deepfakes for 10$, dm me"*, *"I can pay, we need only 30sec or 1min deepfake video"* in the subreddit SFWdeepfakes. These actions give cause for concerns over the moral intuition of the communities on such platforms. Moral intuition is judgmental and based on human behaviors; thus, capturing and examining one's moral intuitions are challenging, due to linguistic

and behavioral complexities. Nevertheless, due to the potentially harmful nature of the activities resulting from deepfakes, it is important to examine how communities perceive deepfakes and whether the emerging behavior is morally acceptable to society. The term "morality" is identified as the judgment of doing the right or wrong thing in society's eyes (Capraro and Rand, 2018); every individual has moral values that play a key role in their behavior development (Filip et al., 2016).

Regardless of any particular ethical framework that one holds, the common practice of evaluating others' moral behavior is widely regarded as important for the well-being of a community. Despite the complexities involved in capturing moral intuitions, ethnographers and sociologists have attempted to develop empirical understandings of how moral intuitions change across contexts—e.g., platforms where moral judgments increase cooperation within a community (Boyd and Richerson, 1992), or how moral behaviors divide communities and negatively impact the society (Weaver and Lewis, 2012; Dubljević et al., 2022). Research indicates that increased volumes of sharing behavior on social media and the platforms' abilities to mix viewpoints across platforms inevitably lead to online debates around moral judgment on various topics—particularly political and social issues, such as gun violence, immigration rights, and governmental policy support (Horberg et al., 2011), or cultural issues, such as meat consumption (Bryant and van der Weele, 2021), polygamy, or child marriage (Hooker, 2003), are the subject of discussions of what is morally acceptable or unacceptable. The content of these debates provides researchers with the opportunity to ask specific questions about an argument, disagreement, moral evaluation, and/or judgment with the aid of new computational tools. Similar to topics that involve extensive moral debates, trending technologies such as deepfakes may offer debatable outcomes and may also lead to harmful outcomes to society, yet these may not be clearly visible in community discussions. The intuitions behind discussions on deepfakes may pose many questions on its morality; however, we have yet to understand how individual actions, institutions, and communities react to the use of deepfakes, nor how these users perceive deepfakes in terms of their moral intuitions. On many occasions, deepfakes have been used for the purpose of deliberate manipulation, tarnishing reputations, or even threatening or blackmailing purposes (Pantserev, 2020). Irrespective of the applications of deepfakes, the moral intuitions behind the public discussions of deepfakes can lead to social outrage and can impact society at large. It has been five years since the r/deepfakes ban was implemented (Brooks, 2021), yet we still see many activities on Reddit about deepfakes. This poses the question of whether these activities can lead to social harm.

Since potential social harm can be identified by examining the intuitions behind the conversations that potentially lead to immoral behaviors, we examined the moral intuitions behind Reddit discussions concerning deepfakes. Specifically, we conducted a text content analysis of discussions of deepfakes from 2018 to 2022. Understanding moral intuition from discussions or text narratives is complex; thus, in this research, we leverage Moral Foundations Theory (MFT) (Graham et al., 2013) to capture the moral intuitions behind the discussions. MFT offers a framework for identifying moral intuitions in five key dimensions: Care (or Harm); Fairness (or Cheating); Loyalty (or Betrayal); Authority (or Subversion); and Sanctity (or Degradation). We use eMFDscore, an Natural Language Processing (NLP) tool that uses an extended Moral Foundations Dictionary (eMFD) to quantify the moral loadings (as a proxy for moral intuitions) behind the discussions (Hopp et al., 2021). We demonstrate the different strengths of moral intuitions and the evidence of harmful conversations rooted in these intuitions. Although the degree of strengths and the endorsement of these five moral dimensions vary from one individual to another (Lifton, 1985), stronger moral dimensions collectively shape and normalize certain judgments in a community. In contrast, doing right or wrong to society by using deepfakes depends on the moral behavior of individuals. Since the Reddit community is one that is very specific about controversial topics or technologies, their moral judgements may impact the application of deepfakes. We draw empirical judgements as to whether communities morally accept deepfakes and which moral intuitions are reflected in the Reddit community. To comprehend such a holistic community viewpoint, we specifically ask the following two research questions:

*1. What are the strengths of the moral intuitions behind the deepfake conversations on the Reddit platform in 2018–2022?*

*2. How have these moral intuitions have changed over the years?*

## 2 Literature Review
### 2.1 Deepfakes
As part of the advancement of AI technologies, deepfakes have been introduced as a way to swap faces in video and digital content to create realistic-looking fake media (Nguyen et al., 2022). Not only face-swapping images but also synthetically generated videos and audio are widely popular and are used for different purposes. Many applications are emerging that enable creating deepfakes much more easily, including the Chinese app Zao, DeepFace Lab, and FaceApp (Chadha et al., 2021). As technology grows and improves at the current pace, it is expected that deepfake technologies will become much more sophisticated and accessible to many, and they might introduce more serious threats to the public related to election interference, political tensions, and additional criminal activity (Chadha et al., 2021).

The trends of deepfakes threats predicted in research are becoming reality and making headline news: stories include a deepfake video of former president Donald Trump giving a speech calling on Belgium to withdraw from the Paris climate agreement (Burchard, 2018), a mother of teenage daughter using deepfake pornographic videos to threaten her daughter's rivals on a cheerleading squad (Guardian, 2021), hackers use deepfakes videos of an executive at the crypto currency platform Binance to scam multiple crypto projects (Barr, 2022), and more. Although deepfakes technologies can be used for good (Kwok and Koh, 2021)—such as in the film industry, education content creation, or medical field—they pose incredible challenges in terms of security threats, ethical or moral dilemmas, and social outrage.

Deepfakes cannot be labeled as solely problematic, and indeed the technology can be used to produce many positive and useful outcomes, but the dilemma nevertheless involves the negative outcomes it can create (de Ruiter, 2021). Additionally, they exist in a social environment rife with "toxic-technocultures", with questionable moral values and attitudes. The value of the technology is at risk due to the social behaviors aroused in response to the deepfake technology (Massanari, 2017; Kugler and Pace, 2021). In other words, sometimes the issue is not just the technology itself, but the way the community discusses it, embraces it, and acts on it that triggers severe consequences. The Reddit platform design, which is the context of this research, has been criticized around its design, governance, algorithm, and platform politics for implicitly supporting kinds of cultures that are problematic to society (Massanari, 2017). Therefore, we highlight that conversations about deepfakes taking place on social platforms can lead to problematic outcomes. We need to capture the moral intuitions of such conversations for much more effective monitoring and regulating.

Although considerable research has addressed the concerns of deepfakes through reviews (Gamage et al., 2021a), the public perspective (Dobber et al., 2021; Yaqub et al., 2020), and philosophical perspectives (Westling, 2019; Ohman , 2019), limited research has empirically explored how deepfakes can impact society (Gamage et al., 2021b). In order to understand how the public perceives the use of deepfakes, a study surveyed 1512 users and explored people's attitudes toward sharing deepfake news headlines with their friends on social media (Yaqub et al., 2020). Other studies have demonstrated the effects of political attitudes on deepfakes (Dobber et al., 2021; Gamage et al., 2021a). A few other studies have analyzed deepfake content on social media: one analyzed YouTube comments about deepfakes (Lee et al., 2021), and another analyzed journalist discourse to understand the social implications of the deepfakes (Wahl-Jorgensen and Carlson, 2021). However, thus far, the moral intuitions behind deepfakes have not been examined. The most similar studies to our research were an analysis of deepfake-related tweets in terms of sentiments, geographies, and key users who constantly share deepfakes (Dasilva et al., 2021) and an examination of topics in deepfake-related discussions on Reddit (Gamage et al., 2022).

Unlike previous analyses, our focus is on the moral intuitions behind discussions on deepfakes. To this end, we specifically analyzed Reddit data, since Reddit users introduced deepfakes in the first place and the Reddit community has a history of promoting toxic technocultures (Massanari, 2017). However, capturing moral intuitions behind discussions and identifying problematic behaviors are complex, due to the nature of language. Previous studies have proposed a three-factor framework to understand whether deepfakes are morally problematic: investigate whether the deepfaked person(s) would object to the way in which they are represented; investigate whether the deepfake deceives viewers; and investigate the intent with which the deepfake was created (de Ruiter, 2021). However, no tools exist that can operationalize this framework or apply it in the context of a large discussion forum. Indeed, empirical tools that could provide operational mechanisms to detect the harmful moral intuitions of deepfakes are lacking; this is concerning, given the fact that these technologies are accessible to many users.

As a precaution to reduce the harm of deepfakes, many social media platforms have banned deepfakes at many levels (Lucas, 2022), including Reddit, Facebook, Twitter (van der Sloot and Wagensveld, 2022), and the Google platform Colaboratory (Wiggers, 2022). However, we argue that banning deepfake content from platforms is a temporary solution. Many behaviors on social media relating to deepfakes are concerning, yet it is not possible to detect the intuitions behind the use of deepfakes or how the community perceives deepfake use with the existing content policies.

## 2.2 Reddit Communities
Reddit is a social news aggregation, content rating, and discussion website that is composed of many subcommunities, known as subreddits (Singer et al., 2014). Each subreddit has a specific topic they discuss, such as technology, politics, music, etc. (Singer et al., 2014). The platform is founded and centered on the concept of freedom of speech; in the past, it has shown resistance to censoring its users, despite the prominence of racist, misogynistic, homophobic, and explicitly violent material on the platform (Copland, 2020).

However, due to pressure from its users, the general public, and lawmakers, Reddit has begun to censor its content. Specifically, sexual content involving minors or nude pictures, conversations that incite violence or attacks, and harassment of broad social groups led to the ban of certain users or whole subreddits; a list of these subreddit names can be found in Wikipedia (Wik, 2022). Most of these subreddits were banned due to its community's activities featuring graphic depictions of violence against women (r/Beatingwomen), hosting photos of overweight people for the purpose of mockery (r/fatpeoplehate), and superimposing famous female actresses onto pornographic videos (r/deepfakes). The r/deepfakes subreddit was the community that introduced deepfakes to the world (Fletcher, 2018). Although it was banned based on the platform's policy against nonconsensual nude pictures posted in the community, deepfake-related discussions still continue on Reddit in various formats. Therefore, banning and regulating are heavily dependent on platform policies and on community's norms and guidelines (Chandrasekharan et al., 2018).

Recent research has found that nearly 4% of Reddit posts and comments in 2020–2021 violated community guidelines and norms, and that many anti-social behaviors are likely not being moderated. Pornography and bigoted comments were more likely to be moderated, while political inflammatory comments, misogynistic, and vulgar comments were the least likely to be moderated (Park et al., 2022). In terms of abusive behavior, Reddit regulations rely on user-level interventions, such as community moderators, or platform-level interventions, such as quarantining the group or banning the entire subreddit or series of subreddits (Chandrasekharan et al., 2022; Habib et al., 2022). Moderators are primarily volunteers designated by subcommunity administrators who regulate content according to community rules (Chandrasekharan et al., 2019). Users also play a role of content monitors, by reporting content to the platform to initiate moderation (Gillespie, 2018).

Although we see continuous efforts to optimize moderation techniques using proactive and reactive tools (Habib et al., 2022) or human moderators (Chandrasekharan et al., 2017), there is evidence that many hateful comments or violations occur in Reddit communities (Park et al., 2022). In addition, community norms may not always capture certain unethical discussions. How can we better support the community in being aware of such harmful or unethical discussions that not necessarily be immediate visible as harmful but raise what is socially right from wrong, is something that being unexplored. Although much work has been done to understand toxic subreddit behaviors, hate speech (such as that in r/The Donald) (Chandrasekharan et al., 2022), harassment, misogyny (such as that seen in r/MGTOW) (Farrell et al., 2019) or mental health (such as r/depression, r/anxiety) (Sharma and De Choudhury, 2018), to our knowledge, "morality" or the moral intuitions behind deepfakes have not been explored on Reddit.

The closest research we can find to examining moral intuitions behind deepfakes entail an exploration of /r/AmITheAsshole through MFT using BERT classification (Botzer et al., 2022), an evaluation of whether Reddit quarantines for r/The Donald and r/ChapoTrapHouse had an impact on value associations highlighted in political discussion (Shen and Ros´e, 2022), and an exploration of deepfake discussions in terms of key topics (Gamage et al., 2022). These studies provide insights but lack empirical analysis of the moral intuitions behind discussions of deepfakes. Given that everyday users are increasingly engaged in judgments of the technologies they use, deepfakes, and how these technologies (and their ecosystems) interact with a community, can be deemed "ethical" or "unethical" in terms of moral values.

## 2.3 Moral Intuitions

Individuals' moral intuitions play an instrumental role in a social process, specifically when taking decisions and actions. We witness the flux of moral intuitions in occasions such as in voting (Morgan et al., 2010), moralizing vaccine hesitance (Amin et al., 2017), the selection, valuation, and production of media content (Tamborini and Weber, 2020), characters in online gaming (Arrambide et al., 2022), or even in instigation of violent protests (Mooijman et al., 2018).

To understand individual morality, a group of social and cultural psychologists created Moral Foundations Theory, or MFT, which explains the variations in human moral reasoning based on innate, modular foundations (Graham et al., 2013). MFT has been using as an instrument to understand political, religious, and cultural differences between individuals and groups. The instrument helps determine the roots of users' moral beliefs and how social systems contribute to decision-making concerning morally laden topics. Although the authors indicate the flexibility in expanding the dimensions that capture morality, moral intuitions were originally operationalized on five moral dimensions (Graham et al., 2011):

• **Care/Harm** is caring behavior toward other group members who are in need of protection. Harm is the opposite of this behavior, wherein no care is expressed toward others.

• **Fairness/Cheating** is associated with sensitivity to inequality and the motivation to maintain justice within the group. When people are insensitive to injustice, it is considered Cheating.

• **Loyalty/Betrayal** is protecting the interests of one's own group, favoring one's own group member, and discriminating against out-groups. The opposite behavior to this is Betrayal.

• **Authority/ Subversion** is the desire to maintain the hierarchical structure in the group, as well as respect for those who are higher in authority; the opposite of this is identified as Subversion.

• **Sanctity or Purity/degradation** is related to the suppression of desires, a motivation to be pure both physically and spiritually and to avoid infectious diseases. In other words, purity is the antipathy of disgusting behaviors. Its opposite is Degradation.

To understand group dynamics, these five moral foundations are categorized into two superordinates: i) Individualizing foundations, consisting of Care/Harm and Fairness/Cheating, sensitize people to suffering and consider the equitable treatment of individuals; and ii) Binding foundations, consisting of the three other foundations, bind people together as groups (Graham et al., 2013, 2011).

Numerous studies have utilized MFT to examine societal phenomena or individual attitudes. For example, research has attempted to understand whether morality or political ideology determines attitudes toward climate change (Dawson and Tyson, 2012), to explain capital jurors' sentencing decisions based on moral foundations (Vaughan et al., 2019), and to understand how culture shapes moral identity in Britain and Saudi Arabia (AlSheddi et al., 2020). In the context of social media platforms, researchers have used MFT to examine human values and attitudes toward vaccination on social media (Kalimeri et al., 2019), while another study understood cultural differences in language use in Japanese and English by analyzing tweets (Singh et al., 2021). Given that it is a widely applicable and adoptable empirical framework, we use MFT as a basis for our examination of the moral intuitions behind deepfake discussions in the Reddit community.

Although MFT provides a systematic framework to describe morality, measuring and interpreting the social influence of morally relevant communications in communities is challenging (Garten et al., 2018; Weber et al., 2018). When measuring the moral intuitions behind large-scale textual data, researchers need an automated computational method. Existing computer-assisted extraction of moral content relies on lists of individual words, initially compiled in the Moral Foundations Dictionary (MFD) (https://moralfoundations. org/). The MDF has a list of 324 English words capturing each foundation of MFT in terms of its vice or virtue of moral intuitions. Although the number of words in the dictionary has expanded (Rezapour et al., 2019) over time, researchers have argued that the words in the MFD are less representative for measurements of the moral intuitions of human beings (May, 2018). To resolve this, Hopp et al. (Hopp et al., 2021) introduced an extended Moral Foundation Dictionary (eMFD), which comprises 3270 representative words based on crowd-sourced, context-laden annotations of text that indicate moral intuitions. The eMFD outperforms previous moral extraction approaches in a number of different validation tests (Hopp et al., 2021). In this research, we used the eMFD to measure the moral intuitions behind the discussions on deepfakes.

## 3 Methods

### 3.1 Measurement of Moral Loadings

For the dataset, we compute moral loadings using eMFDscore, which is a Python package that calculates the probability of occurrences of moral words from the five moral foundations. The eMFD has more than 3000 words, and each word is assigned five probability values for the likelihood of that word being associated with each of the five moral foundations as identified by MFT. For example, in the case of a sample word "kill", it has a Care/Harm probability of 0.4 and a Loyalty/Betrayal probability of 0.24, meaning that there is a 40% chance that the context in which the word "kill" appeared was highlighted with the Care/Harm foundation and a 24% chance that this context was highlighted with the Loyalty/Betrayal foundation.

To understand whether the morality is virtue or vice, we rely on the sentiment score result in the eMFDscore. Five sentiment scores are assigned to each word, denoting the average sentiment of the foundation context in which this word appeared. For example, the word "kill" has an average "care sent" of -0.69, meaning that this is a negative sentiment and therefore is categorized as vice. A complete explanation of the eMFDscore, the dictionary, sample data, and sample analyses are available at https://github.com/medianeuroscience/ emfd) and https://osf.io/vw85e/ (Hopp et al., 2021).

Using our dataset, we measured moral loadings and sentiment scores for each post from 2018 to 2022. The data frame contained 101,869 posts and comments, organized in chronological order based on the time stamp. When we submitted the frame to eMFDscore, it provided measurements for 101,869 documents, or all posts in our dataset. Each document was given a moral loading value for the five foundations and a sentiment value that was calculated based on the built-in VADER sentiment scoring. We classified moral foundations based on the highest moral loading and the sentiment based on the highest moral foundation estimate in eMFDscore. Figure 1 illustrates the input and output data: the eMFDscore inputs the dataset with text and date, and it outputs five moral loadings based on probability scores, five sentiment scores, the moral to non-moral ratio along with the variance across the foundation scores (f var), and the variance across the sentiment scores (sent var).

Next, based on the highest probability score provided by eMFDscore, we classified each post into one of the five moral foundations. If the sentiment score was negative, we considered the morality as vice; if the sentiment score was positive, we considered the morality as virtue. The resultant moral loadings are plotted as a time series to observe trends and as boxplots to draw comparisons across moral foundations by year, which enables us to understand moral intuitions behind deepfake-related discussions on Reddit. To observe overall years of conversations, we created wordclouds using conversations that were classified to each moral foundation; in the cloud, the size of a word is proportional to its occurrence frequency, and the colors represent different moral foundations. Before we plotted the wordcloud, we removed the most-occurring word ("deepfake") from our dataset, to obtain more visibility on the type(s) of words seen in the moral foundations.

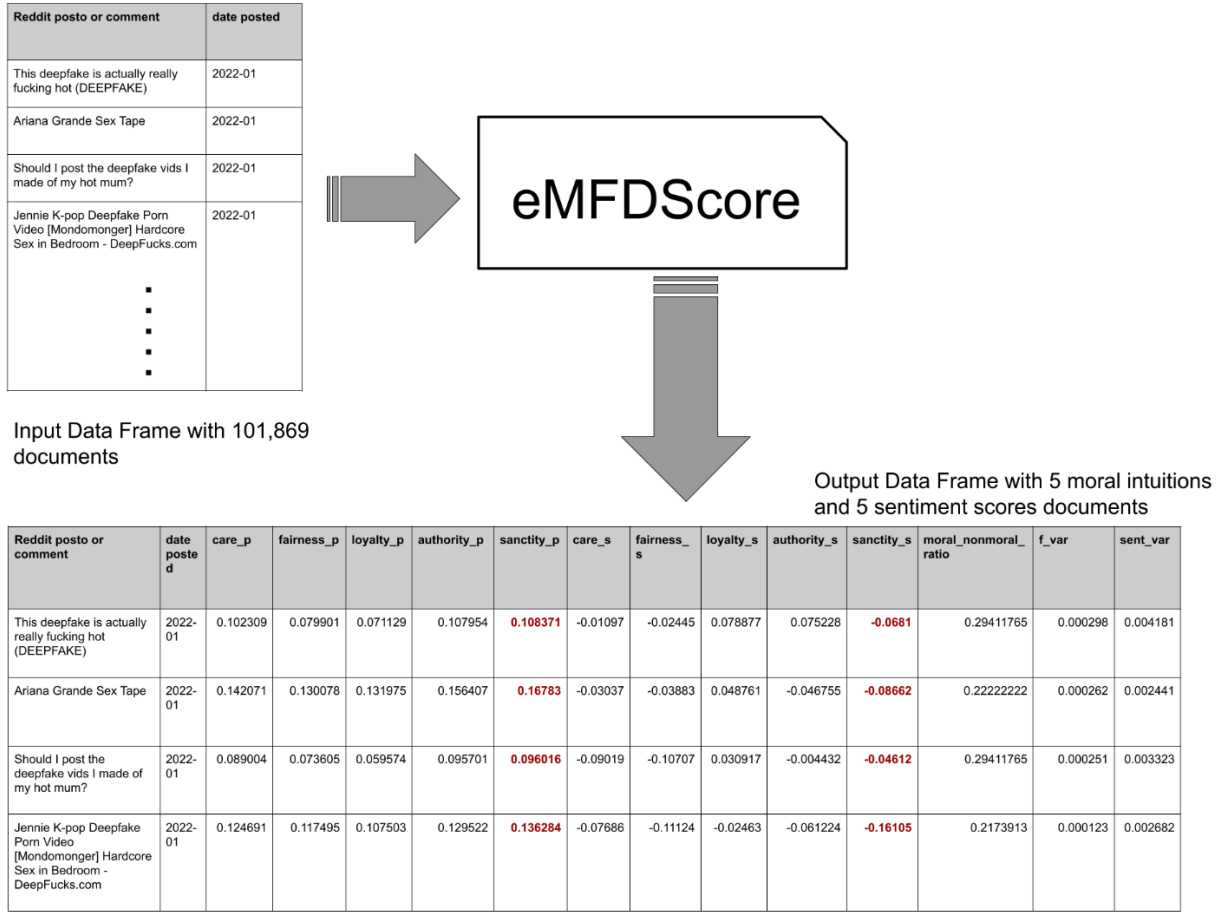

Figure 1: Example of input to and output from the eMFDscore, with the highlight for the highest moral loadings and the relevant sentiment.

When using the eMFDscore, we set the parameters seen in Table 1. However, other options could have been selected as parameters. For example, SCORE METHOD could be set to wordlist scoring algorithm, instead of the method bow that we used. The wordlist method lets users examine the moral content of individual words instead of a 'bag-of-words'. Since our motivation was to extract the overall holistic moral intuition from the text document, we used the recommended bag-of-words method with all probabilities per word. For example, the text of a post in March 2018 (the n = 86645th post) had the text *"We Are Truly Fucked: Everyone Is Making AI-Generated Fake Porn Now. A user-friendly application has resulted in an explosion of convincing face-swap porn. And it started here on reddit"*; the corresponding moral loadings were `care_p` 0.1053, `fairness_p` 0.1197, `loyalty_p` 0.1022, `authority_p` 0.0959, and `sanctity_p` 0.0952. Given the highest probability (0.1197), this post is most likely reflecting the Fairness/Cheating moral dimension. The library also provided the sentiment scores of the post based on the moral words detected in the text. In the above example, the sentiment scores were `care_sent` 0.0224, `fairness_sent` 0.0368, `loyalty_sent` 0.0757, `authority_sent` 0.0210, and `sanctity_sent` -0.0301, which means that the post is reflecting a positive morality in the Fairness moral foundation. The user who posted the statement was concerned about justice, or what happens to the community in case of AI-generated or fake porn.

Table 1: Parameter setting for the eMFDscore

| Parameter | Set value | Description |
| --- | --- | --- |
| DICT_TYPE | eMFD | Use the extended moral foundation dictionary |
| PROB_MAP | All | Use all probabilities per word in the eMFD |
| SCORE_METHOD | Bow | Use Bag-of-Words approach against the dictionary |
| OUT_METRICS | Sentiment | Return the average sentiment for each foundation |
| OUT_CSV PATH | all-sent.csv | Naming the output data frame |

### 3.2 Dataset and Ethical Considerations

Data were collected using the Pushshift Reddit API (https://github.com/ pushshift/api). The search term "deepfake" was used within the time frame January 1, 2018, to September 1, 2022. The data contained posts and relevant comments related to the keyword, subreddit names, usernames of posters and commenters, and date of the post. The script returned a 103MB file with 101,869 submissions (including posts and replies to the posts). We only obtained public conversations, to avoid ethical concerns over collecting Reddit data. We maintained anonymity throughout our research and avoided focusing on usernames or revealing any identities, focusing only on the conversation text by time stamp. Our data and methods are available at OSF (https://osf.io/x5vds/).

## 4 Results

### 4.1 Overall distribution of moral loadings

The overall distribution of moral loadings across the five moral foundations was calculated to determine the most representative moral foundation in deepfake discussions on Reddit. The percentage of each foundation, along with the corresponding mean and standard deviation computed by the eMFDscore, are as follows:

• "Care/Harm" (22%, 0.0982, 0.0105)

• "Loyalty/Betrayal" (21%, 0.0937, 0.0066)

- "Fairness/Cheating" (21%, 0.0954, 0.0080)
- "Authority/Subversion" (20%, 0.0901, 0.0102)
- "Sanctity/Degradation" (16%, 0.0739, 0.0082)

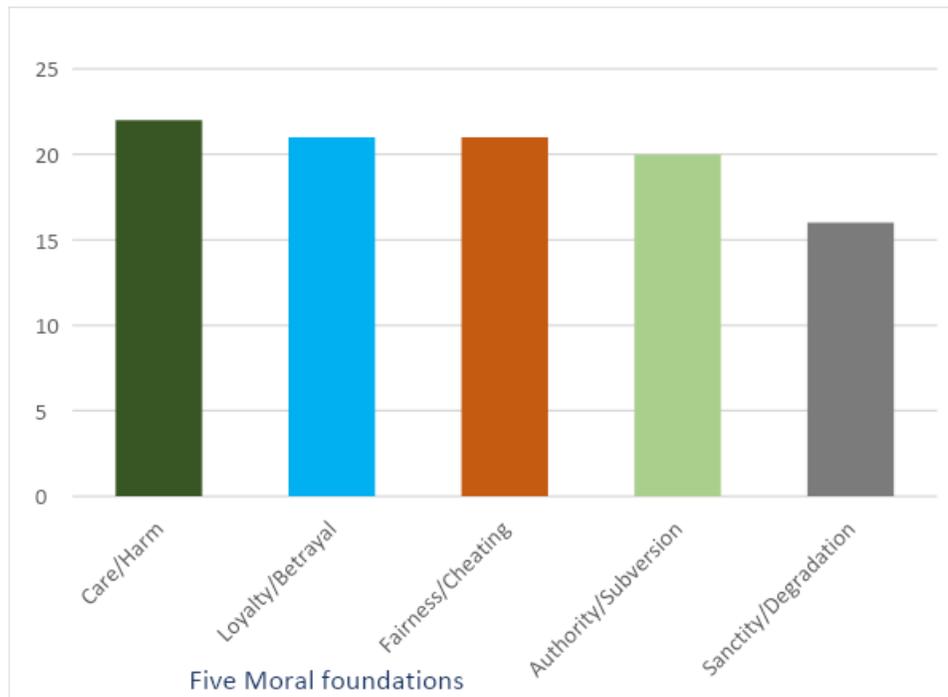

Figure 2: Overall moral loadings 2018–2022 as a percentage

Although the highest and lowest proportions are clearly not visible in Figure 2, most of conversations belonged to the Care/Harm foundation, and the least number of conversations belonged to the Sanctity/Degradation foundation. To understand if the differences between moral foundations are significant, we performed a Friedman nonparametric test, which showed differences between the average moral loadings (Test Statistic = 265121.863 $p$ = 0.000). Subsequently, a post-hoc Dunn test with Bonferroni adjustment showed a significant difference between Care/Harm and Fairness/Cheating ($p$ = 0.000), between Care/Harm and Loyalty/Betrayal ($p$ = 0.00), between Care/Harm and Authority/Subversion ($p$ = 0.00), and between Care/Harm and Sanctity/Degradation ($p$ = 0.000). However, there was no significant difference between Fairness/Cheating and Loyalty/Betrayal ($p$ = 0.135). Thus, in deepfake-related discussions on Reddit, Care/Harm has the highest moral loading, Fairness/Cheating and Loyal/Betrayal have the second-highest loading with no significant difference between the two, and Authority/Subversion and Sanctity/Degradation are third and fourth, respectively, with significant differences between them.

### 4.2 What do higher moral loadings in some foundations mean?

These differences in moral loadings provide important insight into the Reddit community as to their moral intuitions about deepfakes. Their behaviors are more oriented toward *Individualizing foundations*, which means that the community's sensitivity to others or its empathy for others' suffering is reflected more in their posts. This could be either virtue or vice behavior, based on whether the sentiment is positive or negative in the eMFDscore. Nevertheless, Fairness/Cheating and Loyalty/Betrayal have similar strengths: the lack of significant difference between these foundations indicates that the community is also reflecting Binding behavior (i.e., when the community's intuitions are loyalty to each other and respect of ideas). However, the bigger moral question deals with the reason and target of this community's loyalty and care.

To provide more insight into the conversations belonging to the five moral foundations in their virtue and vice, in Table 2 we provide some examples extracted from the eMFDscore. It can be seen that conversations classified as Sanctity/Degradation mostly led to immoral conversations and are inappropriate for any community. Specifically, most of these conversations were related to deepfakes for nude or porn pictures, and the wordcloud in Figure 3 illustrated that the conversations in Sanctity/Degradation are centered on 'edit', 'nude', 'people', 'someone', 'anyone', 'sex', etc. At the same time, in the Care/Harm foundation, the most visible words were 'post', 'think', 'interested', 'got deleted', 'battle', etc., which reflects how the community is protective of the vulnerable. The Fairness/Cheating moral foundation should reflect doing the right thing or justice based on shared rules; however, our dataset did not visibly illustrate whether the community is against unethical or immoral conversations regarding deepfakes. Similarly, in the Authority/Subversion moral foundation, the words that appeared were 'channel', 'created', 'real', 'interested', etc., which did not illustrate a hierarchy or obeying/respecting the tradition but rather reflected initiative actions taken by the community. Sample posts depicted in Table 2 show how a virtue comment in the Authority/Subversion foundation is about a deepfake created and a vice comment is about moderation of the community.

Table 2: Sample posts based on eMFDscore

| Moral Foundation | Valence (Virtue/Vice) | Sample post |
| --- | --- | --- |
| Care/ Harm | Virtue | "As deepfake potential grows, what do you see as a limiting factor for this technology?" |
|  | Vice | "Incredible AI deepfake technology compiles every serial killer into one man." |
| Fairness/ Cheating | Virtue | "IT savy people of reddit, how can a regular person combat deepfake content?" |
|  | Vice | Any Free programs for making a deepfake? |
| Loyalty/Betrayal | Virtue | Anyone want to Deepfake nude your crush or anyone?" |
|  | Vice | "You can do deepfake using images, no need nudes of yours, this can be generated" |
| Authority/ Subversion | Virtue | "Flag of Canada except it's just that deepfake of Sartorius giving head " vice |
|  | Vice | "Could you keep this in low cuz mod " |
| Sanctity/ Degradation | Virtue | *"A picture of my mom in a sexy dres*** I like to deepfake her onto pornstars"* |
| | Vice | *"Can somebody deepfake a pic of my hot friend?"* |

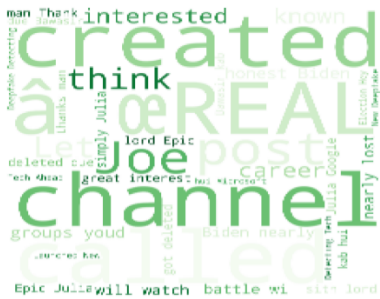
Authority

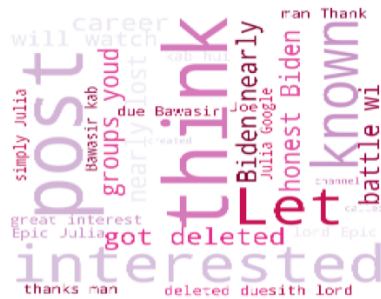
Care

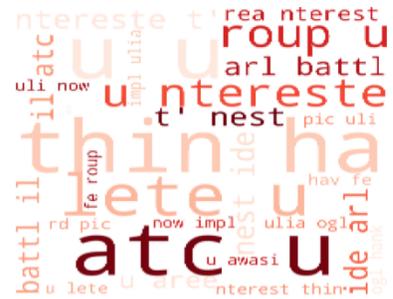
Loyalty

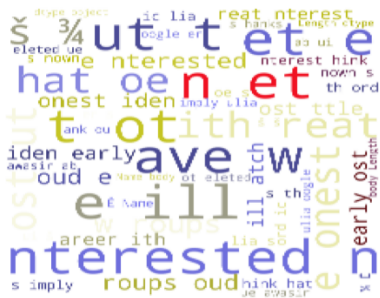
Fairness

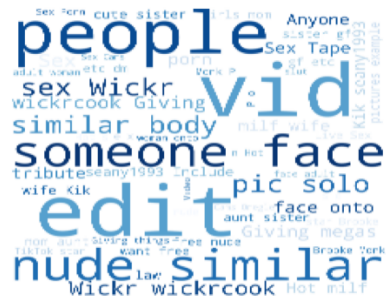
Sanctity

Figure 3: Wordclouds generated from Reddit posts belonging to the five moral foundations

## 4.3 Changes of Moral Intuitions over Time

Next, we examined how each moral foundation varied over the five years of the research period. We plotted a line graph (Figure 4) and a boxplot (Figure 5) based on the values of the five moral foundations resulting from the eMFDscore containing conversations in 2018–2022. To understand the sentiments of these moral foundations, we extracted the sentiment scores (which comprise negative and positive values of each foundation) and plotted them in Figure 6. Using these components, we aim to understand significant patterns in the moral intuitions demonstrated on the Reddit platform over the years—specifically, when the moral intuitions show individualizing values, which promote personal rights and freedoms, and binding values, which govern behavior in groups.

### 4.3.1 Moral intuitions behind individualizing values

We examined how the distribution of the Individualizing foundations Care/Harm and Fairness/Cheating changed over the years. The Care/Harm foundation indicates the communities' sensitivity to each other or their sharing of caring behavior for each other's needs. As described earlier, this foundation has the overall highest moral loading in the Reddit platform. However, to determine if this morality changed over the years or was high for certain years, we performed a nonparametric Kruskal-Wallis test for `care_p` values of posts in the years 2018, 2019, 2020, 2021, and 2022. The results indicate a difference in the expression of Care/Harm over these years (Test Statistic= 20250.603, $p$ = 0.000).

Subsequently, we conducted a Dunn's post-hoc test and found that 2022 was significantly higher than all the other years—2018, 2019, 2020 and 2021 ($p$ = 0.000). Apart from the year 2022, Care/Harm was found to be significantly higher in 2021 than in the other years ($p$ = 0.000). The mean probability of `care_p` in 2019 ($\bar{x}$=0.096) is comparable to that in 2018 ($\bar{x}$=0.096). In Figure 5, it can be seen how this foundation's expression is distributed over the five years. It is worth noting that even though the data for 2022 did not cover the full year, Care/Harm was higher in that year than in any other year. As it gets higher each year, the probability of Care/Harm is trending positively over time.

To check why we see higher Care/Harm in some years than other years, we plotted the frequency of posts based on this moral foundation (see Figure 4). This figure shows some spikes of posts relating to Care/Harm throughout the years 2018–2022. The highest spike occurred in March 2021, when many discussions addressed a series of TikTok videos of a deepfake Tom Cruise and other types of deepfake news shared on the platform. The second highest spike was in May 2019, discussing information about the Chinese government regulating AI face swaps. In terms of the sentiments of this moral foundation, we found that discussion posts leaned toward positive sentiments, which were highest in 2022. In other words, discussions related to this moral foundation reflected care for each other's needs. This positive sentiment of support for each other in conversations about deepfakes on this platform is a unique quality that we found throughout the years.

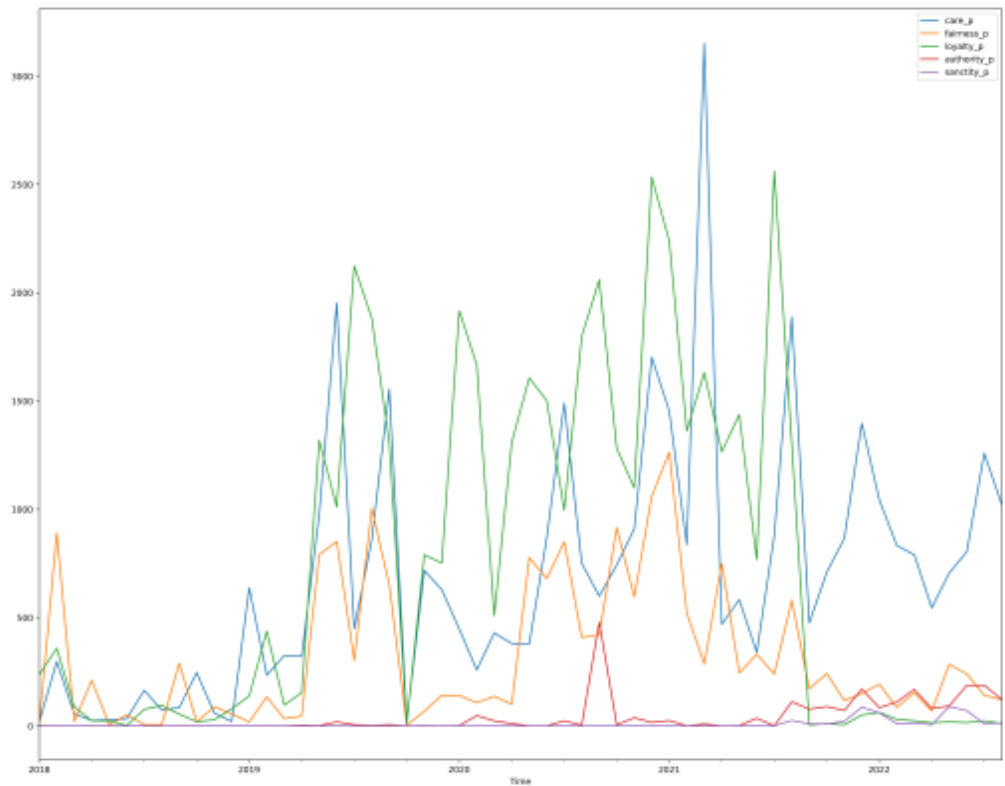

Figure 4: Based on the highest moral loadings from eMFDscore, the frequencies of moral loadings were plotted as a function of time (2018–2022). Care/Harm was found to be the most common moral foundation. Loyalty/Betrayal and Fairness/Cheating were found to be the next most common, and Authority/Subversion and Sanctity/Degradation were found to be the least common.

The Fairness or Cheating moral foundation arises when the community has to justify or act out justice to the group. This is the second-largest moral foundation demonstrated in our study. Whether the community leans toward trust and social justice for each year is reflected in its positive or negative sentiment. We performed the Kruskal-Wallis test against `Fairness_p` to determine if there was a significant difference between years (Test Statistic= 20153.433, $p$ = 0.000). The highest Fairness/Cheating behavior was found in 2022; in 2021, it was not as frequent as previous years, and it was least frequent in 2020. The post-hoc Dunn's test confirmed that 2020 was significantly lower in Fairness/Cheating than 2018, 2019, and 2021 ($p$ = 0.00). The frequency of this moral foundation was higher in 2018 than in 2019 and 2020 ($p$ = 0.00).

To understand why Fairness/Cheating trended down from 2018 to 2020 but regained in 2021 and increased in 2022, we observed the text of the posts in these years along with the events that occurred in these periods. As previously stated, Fairness enables the group to justify its behaviors or speak out against injustice. Interestingly, in 2018, the r/deepfakes subreddit was banned and site-wide policies against nonconsensual nude pictures were emphasized. These were the times when conversations revolved more around the injustice that could be caused by deepfakes—i.e., posting news about deepfakes or fair thought against harmful deepfakes, such as "*Do you think Deepfake technology will make video evidence meaningless in court?*", "*How would you react if one of your friends made a scat deepfake video of you, but didn't share it?*". All these conversations reflected positive sentiments.

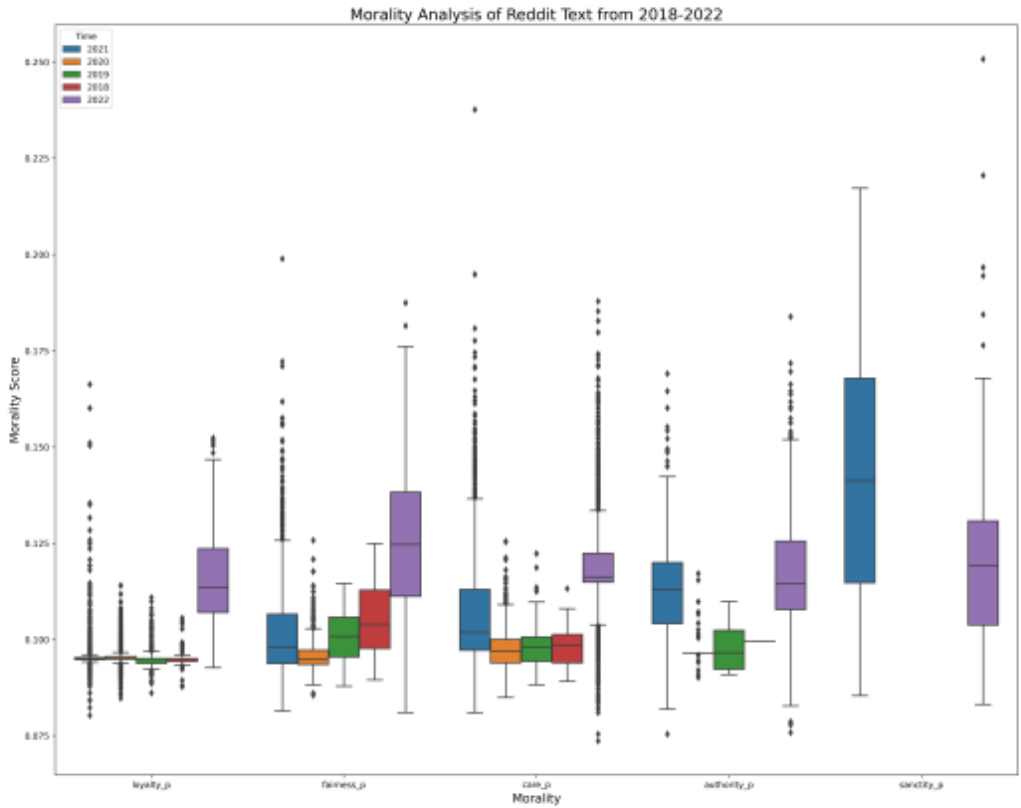

Figure 5: Boxplots were created based on the probability loaded to each post from eMFDscore, grouping the moral foundation and the year. To show the latest effect, 2022 boxplot is in purple; 2018 is in red, 2019 is in green, 2020 is in orange, and 2021 is in blue. There were no occurrences of any posts relating to Sanctity/Degradation during 2018, 2019, and 2020.

However, the trend toward speaking out against injustice about deepfakes ebbed, and from 2021 we observe many discussions about Trump's deepfake videos and many viral posts about other videos that trended. Deepfake video creators are being hired, and since then the conversations have focused less on injustice than on the actions. In 2022, most conversations reflected negative sentiments, which is the opposite of a group acting for social justice in the community. We found that most of the conversations focused more on hiring and providing deepfakes as service, and the community seemed fairly supportive of each other—i.e., "*Deepfake bot three free deepfakes on telegram please use my code so I can get more coins*", "*[reddit] pls dear reddit, does ANYBODY know how to make a deepfake of my mum? i saw some tutorials but they all didn't work for me, can anybody teach or make one for me?*", "*Does anyone know how to deepfake a vdo for free? DM me if you have a way* ", etc.

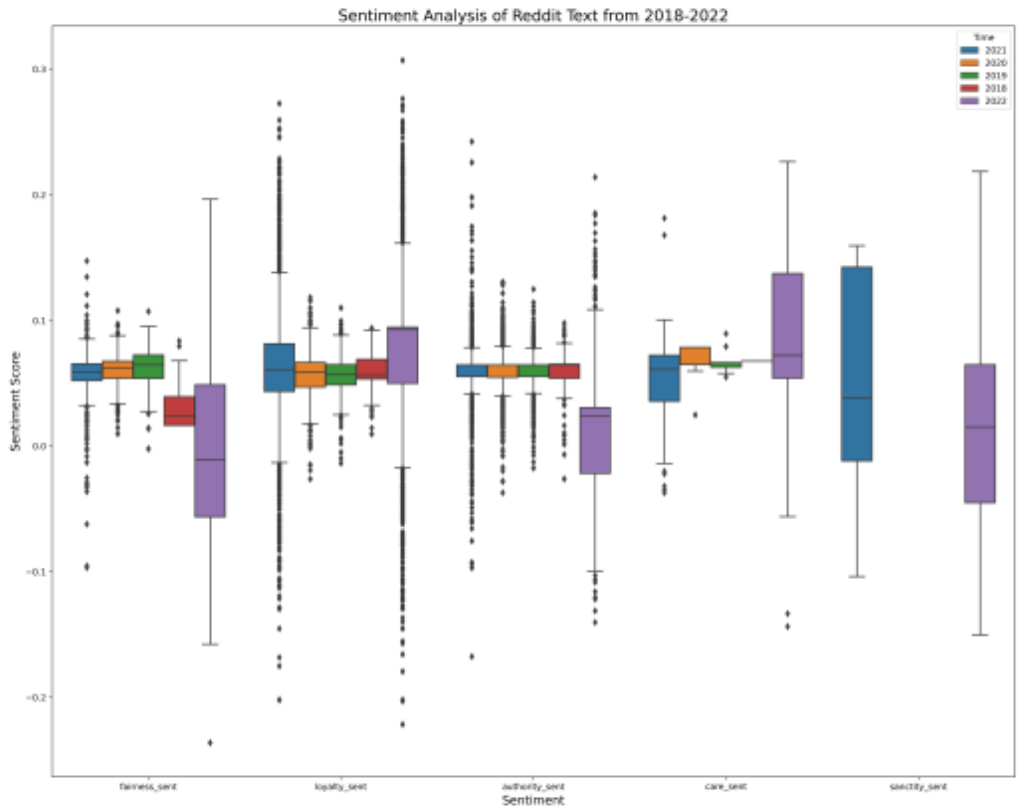

Figure 6: Similar to the strengths of moral foundations loaded to each post from eMFDscore, the boxplots were created based on the sentiment score loaded in each post for 2018–2022. The year 2022 boxplot is in purple, and 2018 is in red, 2019 is in green, 2020 is in orange, and 2021 is in blue. The occurrences of the moral foundation of Sanctity/Degradation in 2021 and 2022 were found to lean toward negative sentiments, reflecting that most posts exhibited Degradation behaviors.

4.3.2 Moral intuitions behind binding values in the Reddit community

Moral intuitions behind the Binding behaviors of the community can be found by examining the conversations of the Loyalty/Betrayal, Authority/Subversion, and Sanctity/Degradation moral foundations. The Loyalty/Betrayal foundation was found to be the second largest foundation overall. As reflected in the name, this foundation involves how a group or community behaves toward each other—in this case, how the Reddit community is loyal to each other in their thoughts and actions. Although the Kruskal-Wallis test against Loyalty_p showed a significant difference between the years (Test Statistic= 21053.433, $p$ = 0.000), the posthoc Dunn's test showed no differences in Loyalty/ Betrayal between 2018, 2019, 2020, and 2021 ($p$ > 0.05). However, interestingly, 2022 showed the highest loyalty behavior, with a significant difference from all the other years ($p$ = 0.000).

This finding is clear evidence that communities in Reddit are more loyal to each other—in our case, loyal to their behaviors concerning deepfakes in terms of positive sentiment. We mostly witnessed creating deepfakes, sharing news, sharing deepfakes, and helping each other with any activities relating to deepfakes. This occurred equally throughout all years and shot much higher in 2022. This trend is an indication that Reddit has not noticed the trending behavior favoring deepfakes; thus, the platform's only policy reining in deepfakes ( non-consensual pornography or deepfakes of minors) seems outdated, as the community has found many other ways to commit actions that are morally questionable—i.e., "*Can anyone deepfake a nude pic for me?*", "*I need to take breast out of this prnbaby,help*".

The Authority/Subversion moral foundation reflects the behavior of communities maintaining a hierarchical structure and respecting those who have authority. Although this foundation is not as dominant on the Reddit platform as Care/Harm and Fairness/Cheating, its distribution over the 5 years was found to significantly vary (Kruskal-Wallis on `Authority_p` Test Statistic= 21253.351, $p$ = 0.000). Although there were no differences in 2018, 2019, and 2020 ($p$ > 0.005), the moral foundation was highly expressed in 2021 and 2022. Examining the posts relating to Authority/Subversion, we found posts relating to commanding or expressions such as "*can you*", "*who can*", "*join me*", "*my work*", "*could you*", *do me…*", etc., mostly serving the community and their needs.

The least frequent moral dimension was Sanctity/Degradation, which is the community's intuition concerning immoral activities or the psychology of disgust and contamination. Most importantly, we found that this foundation only occurs in 2021 and 2022, but not in 2018, 2019, or 2020. Although 2021 was significantly higher than 2022 (post-hoc Dunn's test $p\ value$=0.000), the year 2022 was incomplete, as our data collection only occurred until September 1, 2022; therefore, we believe this could continue increasing through the rest of the year. The sentiment distributed in this moral foundation reflected overall negative distributions. These negative sentiments solely reflected the highly concerning intuitions behind this moral foundation, as each of the posts we observed contained extremely inappropriate actions or thought that render the use of deepfakes questionable in moral terms. Such discussions were centered on vulgar, offensive, and psychologically distressing sexual actions or pornography-related conversations—i.e., "*Find the hottest emma watson sex videos and clips, realy hight quality starletto porn with Russian translation,sex, porno Emma blowjob harry potter deepfake Emma sex porno cosplay*", "*TIFU - my girlfriend left me because she discovered that I'd deepfaked Zoe Kravitz into our sex tapes*".

## 5 Discussion

Extracting moral intuitions is critical to developing an understanding of how human moral behavior and communication unfold at both small and large scales, especially in digital contexts. In this research, we demonstrated how Reddit users act as moral agents in the community, where the strengths of their moral intuitions embedded in their discussions reflect the community's moral behavior toward deepfakes. We evaluated the moral strengths by evaluating users' conversations using eMFDscore. This opened up new dimensions of understanding the communities on Reddit in terms of their moral and immoral behavior.

The contribution of our study is its provision of empirical evidence on the most visible moral intuitions behind the discussions related to deepfakes and how they have trended over the years. We demonstrated the moral intuitions behind discussions classified into five moral foundations. By doing this, we discovered that intuitions under the Sanctity/Degradation moral foundation populating the Reddit platform contained extreme, immoral, and psychologically distressing content. This content should have been flagged or moderated, as it reflects immoral behavior. It seems that platforms do not have mechanisms to moderate or filter immoral conversations, especially in the case of the social system of Reddit. Our research found that some discussions are not against the community rules nor violate any Reddit policies, yet they are morally questionable. The moral intuition results of our study provide evidence underscoring the necessity of big technology platforms (such as Reddit) setting moral obligations to impose ethics not just on deepfakes but on how users engage with discussions relating to deepfakes.

## 6 Limitations and Future Direction

Our analysis utilized a dictionary-based text analysis with eMFDscore to quantify moral intuition in Reddit posts and comments. We used the eMFD, which was created employing a crowd-sourced method, and we used the bag-of-words approach to calculate the probability of each word found according to the eMFD dictionary. This method is intuitively understandable but sometimes less accurate in classification tasks than machine learning (ML) methods, such as pre-trained BERT models on morality (Trager et al., 2022). In the future, classifications of moral intuition can incorporate eMFD words and a set of classified outcomes from relevant sample data to improve both representation and accuracy.

Additionally, as previous research explains, the majority of moral or immoral acts involve an entity engaging in the moral or immoral act and an entity serving as the target of the moral or immoral behavior (Gray and Wegner, 2009), which is described as a moral agent and target relationship. Immoral or moral actions and behaviors of such entities can be examined using the syntactic dependency parsing (SDP) algorithm instead of the bag-of-words algorithm option in eMFDscore (see Table 1, which indicates the parameter SCORE METHOD). Using this small parameter change, one could specifically examine behaviors of specific users in terms of their immoral actions using deepfakes.

## 7 Conclusions

Our aim was to understand the moral intuitions behind conversations around deepfakes on Reddit. We used the MFT to understand moral intuitions and conducted our research using NLP techniques with the eMFDscore. We found that the greatest moral intuition behind deepfake-related conversations was Care/Harm and the least was Sanctity/Degradation. The characteristics of the strengths of moral intuitions reflected important behaviors in this community: in particular, having the greatest Care/Harm intuitions in positive sentiments reflects that Reddit community is bridged and bonded to each other, helping and supporting each other in their deepfake activities. At the same time, and most importantly, the intuitions of the Sanity/Degradation in negative sentiment were found to be highly concerning and disturbing, and they can be classified as morally and ethically unacceptable.

Our work demonstrated that MFT can work as a 'prism' to break down the moral intuitions behind the use of deepfakes on Reddit communities into five interpretable foundations. In order to detect weak signals of emergent deepfake-related threats for the purpose of proactive treatment, we need an advanced contextual understanding of the moral intuitions behind discussions in platforms, one that should be able to identify and classify immoral conversations as early as possible and thus can be used as a prevention technique of societal harm caused by deepfakes.

# References

Al Sheddi M, Russell S, Hegarty P (2020) How does culture shape our moral identity? moral foundations in saudi arabia and britain. European Journal of Social Psychology 50(1):97–110

Amin AB, Bednarczyk RA, Ray CE, et al (2017) Association of moral values with vaccine hesitancy. Nature Human Behaviour 1(12):873–880

Arrambide K, Yoon J, MacArthur C, et al (2022) "i don't want to shoot the android": Players translate real-life moral intuitions to in-game decisions in detroit: Become human. In: CHI Conference on Human Factors in Computing Systems, pp 1–15

Barr K (2022) Hackers use deepfakes of binance exec to scam multiple crypto projects. URL https://gizmodo.com/crypto-binance-deepfakes-1849447018

Botzer N, Gu S, Weninger T (2022) Analysis of moral judgment on reddit. IEEE Transactions on Computational Social Systems

Boyd R, Richerson PJ (1992) Punishment allows the evolution of cooperation (or anything else) in sizable groups. Ethology and sociobiology 13(3):171–195

Brooks CF (2021) Popular discourse around deepfakes and the interdisciplinary challenge of fake video distribution. Cyberpsychology, Behavior, and Social Networking 24(3):159–163

Bryant CJ, van der Weele C (2021) The farmers' dilemma: Meat, means, and morality. Appetite 167:105,605

Burchard HVD (2018) Belgian socialist party circulates 'deep fake' donald trump video. URL https://www.politico.eu/article/spa-donald-trump-belgium-paris-climate-agreement-belgian-socialist-party-circulates-

Capraro V, Rand DG (2018) Do the right thing: Experimental evidence that preferences for moral behavior, rather than equity or efficiency per se, drive human prosociality. Forthcoming in Judgment and Decision Making

Chadha A, Kumar V, Kashyap S, et al (2021) Deepfake: An overview. In: Proceedings of Second International Conference on Computing, Communications, and Cyber-Security, Springer, pp 557–566

Chandrasekharan E, Samory M, Srinivasan A, et al (2017) The bag of communities: Identifying abusive behavior online with preexisting internet data. In: Proceedings of the 2017 CHI Conference on Human Factors in Computing Systems, pp 3175–3187

Chandrasekharan E, Samory M, Jhaver S, et al (2018) The internet's hidden rules: An empirical study of reddit norm violations at micro, meso, and macro scales. Proceedings of the ACM on Human-Computer Interaction 2(CSCW):1–25

Chandrasekharan E, Gandhi C, Mustelier MW, et al (2019) Crossmod: A crosscommunity learning-based system to assist reddit moderators. Proceedings of the ACM on human-computer interaction 3(CSCW):1–30

Chandrasekharan E, Jhaver S, Bruckman A, et al (2022) Quarantined! examining the effects of a community-wide moderation intervention on reddit. ACM Transactions on Computer-Human Interaction (TOCHI) 29(4):1–26

Copland S (2020) Reddit quarantined: Can changing platform affordances reduce hateful material online? Internet Policy Review 9(4):1–26

Dasilva JP, Ayerdi KM, Galdospin TM, et al (2021) Deepfakes on twitter: Which actors control their spread? Media and Communication 9(1):301–312

Dawson SL, Tyson GA (2012) Will morality or political ideology determine attitudes to climate change. Australian Community Psychologist 24(2):8–25

Dobber T, Metoui N, Trilling D, et al (2021) Do (microtargeted) deepfakes have real effects on political attitudes? The International Journal of Press/Politics 26(1):69–91

Dubljevi´c V, Douglas S, Milojevich J, et al (2022) Moral and social ramifications of autonomous vehicles: a qualitative study of the perceptions of professional drivers. Behaviour & Information Technology pp 1–8

Farrell T, Fernandez M, Novotny J, et al (2019) Exploring misogyny across the manosphere in reddit. In: Proceedings of the 10th ACM Conference on Web Science, pp 87–96

Filip I, Saheba N, Wick B, et al (2016) Morality and ethical theories in the context of human behavior. Ethics & Medicine 32(2):83

Fletcher J (2018) Deepfakes, artificial intelligence, and some kind of dystopia: The new faces of online post-fact performance. Theatre Journal 70(4):455– 471

Gamage D, Chen J, Sasahara K (2021a) Emergence of deepfakes and its societal implications: A systematic review. In: Truth and Trust International Conference

Gamage D, Ghasiya P, Bonagiri V, et al (2022) Are deepfakes concerning? analyzing conversations of deepfakes on reddit and exploring societal implications. In: CHI Conference on Human Factors in Computing Systems, pp 1–19

Garten J, Hoover J, Johnson KM, et al (2018) Dictionaries and distributions: Combining expert knowledge and large scale textual data content analysis. Behavior research methods 50(1):344–361

Gillespie T (2018) Custodians of the Internet: Platforms, content moderation, and the hidden decisions that shape social media. Yale University Press

Graham J, Nosek BA, Haidt J, et al (2011) Mapping the moral domain. Journal of personality and social psychology 101(2):36

Graham J, Haidt J, Koleva S, et al (2013) Moral foundations theory: The pragmatic validity of moral pluralism. In: Advances in experimental social psychology, vol 47. Elsevier, p 55–130

Gray K, Wegner DM (2009) Moral typecasting: divergent perceptions of moral agents and moral patients. Journal of personality and social psychology 96(3):505

Guardian T (2021) Mother charged with deepfake plot against daughter's cheerleading rivals. URL https://www.theguardian.com/us-news/2021/mar/15/mother-charged-deepfake-plot-cheerleading-rivals

Habib H, Musa MB, Zaffar MF, et al (2022) Are proactive interventions for reddit communities feasible? In: Proceedings of the International AAAI Conference on Web and Social Media, pp 264–274

Hooker B (2003) Ethics in conflict: Ethical systems. In: Science and Technology Ethics. Routledge, p 97–114

Hopp FR, Fisher JT, Cornell D, et al (2021) The extended moral foundations dictionary (emfd): Development and applications of a crowd-sourced approach to extracting moral intuitions from text. Behavior research methods 53(1):232–246

Horberg EJ, Oveis C, Keltner D (2011) Emotions as moral amplifiers: An appraisal tendency approach to the influences of distinct emotions upon moral judgment. Emotion Review 3(3):237–244

Kalimeri K, G. Beir´o M, Urbinati A, et al (2019) Human values and attitudes towards vaccination in social media. In: Companion Proceedings of The 2019 World Wide Web Conference, pp 248–254

Kugler MB, Pace C (2021) Deepfake privacy: Attitudes and regulation. Northwestern Public Law Research Paper (21-04)

Kwok AO, Koh SG (2021) Deepfake: a social construction of technology perspective. Current Issues in Tourism 24(13):1798–1802

Lee Y, Huang KT, Blom R, et al (2021) To believe or not to believe: framing analysis of content and audience response of top 10 deepfake videos on youtube. Cyberpsychology, Behavior, and Social Networking 24(3):153–158

Lifton PD (1985) Individual differences in moral development: The relation of sex, gender, and personality to morality. Journal of personality 53(2):306–334

Lucas KT (2022) Deepfakes and domestic violence: perpetrating intimate partner abuse using video technology. Victims & Offenders 17(5):647–659

Massanari A (2017) # gamergate and the fappening: How reddit's algorithm, governance, and culture support toxic technocultures. New media & society 19(3):329–34

May J (2018) Regard for reason in the moral mind. Oxford University Press

Mooijman M, Hoover J, Lin Y, et al (2018) Moralization in social networks and the emergence of violence during protests. Nature human behaviour 2(6):389–396

Morgan GS, Skitka LJ, Wisneski DC (2010) Moral and religious convictions and intentions to vote in the 2008 presidential election. Analyses of Social Issues and Public Policy 10(1):307–320

van der Nagel E (2020) Verifying images: Deepfakes, control, and consent. Porn Studies 7(4):424–429

Nguyen TT, Nguyen QVH, Nguyen DT, et al (2022) Deep learning for deepfakes creation and detection: A survey. Computer Vision and Image Understanding 223:103,525

Ohman C (2019) Introducing the pervert's dilemma: a contribution to the ¨ critique of deepfake pornography. Ethics and Information Technology pp 1–8

Pantserev KA (2020) The malicious use of ai-based deepfake technology as the new threat to psychological security and political stability. In: Cyber defence in the age of AI, smart societies and augmented humanity. Springer, p 37–55

Park JS, Seering J, Bernstein MS (2022) Measuring the prevalence of antisocial behavior in online communities. CSCW

Rezapour R, Shah SH, Diesner J (2019) Enhancing the measurement of social effects by capturing morality. In: Proceedings of the tenth workshop on computational approaches to subjectivity, sentiment and social media analysis, pp 35–45

de Ruiter A (2021) The distinct wrong of deepfakes. Philosophy & Technology pp 1–22

Sharma E, De Choudhury M (2018) Mental health support and its relationship to linguistic accommodation in online communities. In: Proceedings of the 2018 CHI conference on human factors in computing systems, pp 1–13

Shen Q, Rosé CP (2022) A tale of two subreddits: Measuring the impacts of quarantines on political engagement on reddit. In: Proceedings of the International AAAI Conference on Web and Social Media, pp 932–943

Singer P, Flöck F, Meinhart C, et al (2014) Evolution of reddit: from the front page of the internet to a self-referential community? In: Proceedings of the 23rd international conference on world wide web, pp 517–522

Singh M, Kaur R, Matsuo A, et al (2021) Morality-based assertion and homophily on social media: A cultural comparison between english and japanese languages. arXiv preprint arXiv:210810643

van der Sloot B, Wagensveld Y (2022) Deepfakes: regulatory challenges for the synthetic society. Computer Law & Security Review 46:105,716

Tamborini R, Weber R (2020) Advancing the model of intuitive morality and exemplars. In: The handbook of communication science and biology. Routledge, p 456–469

Trager J, Ziabari AS, Davani AM, et al (2022) The moral foundations reddit corpus. arXiv preprint arXiv:220805545

Vaughan TJ, Bell Holleran L, Silver JR (2019) Applying moral foundations theory to the explanation of capital jurors' sentencing decisions. Justice Quarterly 36(7):1176–1205

Wahl-Jorgensen K, Carlson M (2021) Conjecturing fearful futures: Journalistic discourses on deepfakes. Journalism Practice pp 1–18

Weaver AJ, Lewis N (2012) Mirrored morality: An exploration of moral choice in video games. Cyberpsychology, Behavior, and Social Networking 15(11):610–614

Weber R, Mangus JM, Huskey R, et al (2018) Extracting latent moral information from text narratives: Relevance, challenges, and solutions. Communication Methods and Measures 12(2-3):119–139

Westling J (2019) Are deep fakes a shallow concern? a critical analysis of the likely societal reaction to deep fakes. A Critical Analysis of the Likely Societal Reaction to Deep Fakes (July 24, 2019)

Wiggers K (2022) Google bans deepfake-generating ai from colab. URL https://techcrunch.com/2022/06/01/2328459/

Wikipedia, Controversial reddit communities (2020). URL https://en.wikipedia.org/wiki/

Yaqub W, Kakhidze O, Brockman ML, et al (2020) Effects of credibility indicators on social media news sharing intent. In: Proceedings of the 2020 chi conference on human factors in computing systems, pp 1–14